\documentstyle[prl,aps,epsf,twocolumn]{revtex}

\newcommand{\la}{\langle}
\newcommand{\ra}{\rangle}
\newcommand{\beq}{\begin{eqnarray}}
\newcommand{\eeq}{\end{eqnarray}}

\newcommand{\btem}{\bibitem}
\newcommand{\vpi}{\vec{\pi}}
\newcommand{\bep}{{1 \over \bar{\varepsilon}}}

\newcommand{\ms}{m_{0\sigma}}

\newcommand{\mev}{\mbox{MeV}}
\renewcommand{\mp}{m_{0\pi}}
\renewcommand{\d}{\partial}

\begin{document}


\draft

\title{
 Optimized Perturbation Theory at Finite Temperature\thanks{
 Talk presented at ``Thermal Field
 Theories and Their Applications'', 
 (Regensburg, Germany, August 10-14, 1998)}\\
 --- Renormalization and Nambu-Goldstone theorem in $O(N)$ $\phi^4$ Theory ---
} 

\author{S. Chiku$^{a}$ and T. Hatsuda$^{b}$}
\address{$^a$ Institute of Physics, University of Tsukuba,
 Tsukuba, Ibaraki 305, Japan}
\address{$^a$ Yukawa Institute for Theoretical Physics,
Kyoto University, Kyoto 606, Japan}
\address{$^b$ Physics Department,
Kyoto University, Kyoto 606, Japan}

\date{\today}
 
\maketitle

\begin{abstract}

 The optimized perturbation theory (OPT) at finite temperature
 ($T$) recently developed by the present authors is reviewed 
 by using $O(N)$ $\phi^4$ theory with spontaneous symmetry breaking.
 The method resums automatically higher loops (including 
 the hard thermal loops) at high $T$ and simultaneously cures 
 the problem of tachyonic poles at relatively low $T$.
 We prove that (i) the renormalization of the ultra-violet
 divergences can be carried out systematically in any given order of OPT,
 and (ii) the Nambu-Goldstone theorem is satisfied for arbitrary $N$ 
 and for any given order of OPT.

\end{abstract}

\section{Introduction}

 Naive perturbation theory is known to break down at finite temperature ($T$).
 The two reasons are the existence of hard thermal loops (HTL) 
 at high $T$ \cite{BP} and the emergence of  
 tachyonic poles at relatively low $T$ \cite{WEIN74}.
 If one adopts self-consistent resummation methods,
 HTL can be summed and tachyonic poles can be removed.
 However, most of the self-consistent methods proposed so far 
 have difficulties of  renormalization \cite{renorm} and/or 
 the violation of the Nambu-Goldstone theorem \cite{okog} at finite $T$. 

 In this talk, we  show that a new loop-wise expansion 
 at finite $T$ recently developed by the present 
 authors \cite{our} can solve these problems.
 
 Our starting point is the optimized perturbation theory (OPT)
 which is a generalization of the mean-field method \cite{NJL}
 and is known to work in various quantum systems \cite{OPT}.
 Its application to field theory at finite $T$ has been 
 first considered by Okopi\'{n}ska \cite{oko} and Banerjee 
 and Mallik \cite{BM}. We further develop the idea and prove 
 the renormalizability and the Nambu-Goldstone (NG) theorem 
 in $O(N)$ $\phi^4$ theory at finite $T$ order by order in OPT.

 The organization of this talk is as follows.
 In section II, we introduce a loop-wise expansion on the basis of OPT. 
 The renormalization of UV divergences and the realization of the
 NG theorem in this method are also discussed.
 In section III, OPT is applied for the $O(4)$ $\sigma$ model 
 which is a low energy effective theory of QCD.
 Summary and concluding remarks are given in  Section IV. 

\section{Optimized Perturbation at $T \neq 0$}
\label{RPT}
\subsection{Hard thermal loops and tachyonic poles}
\label{N-RT}

\begin{figure}[h]
\centerline{
    \epsfxsize=8.9cm 
    \epsfbox{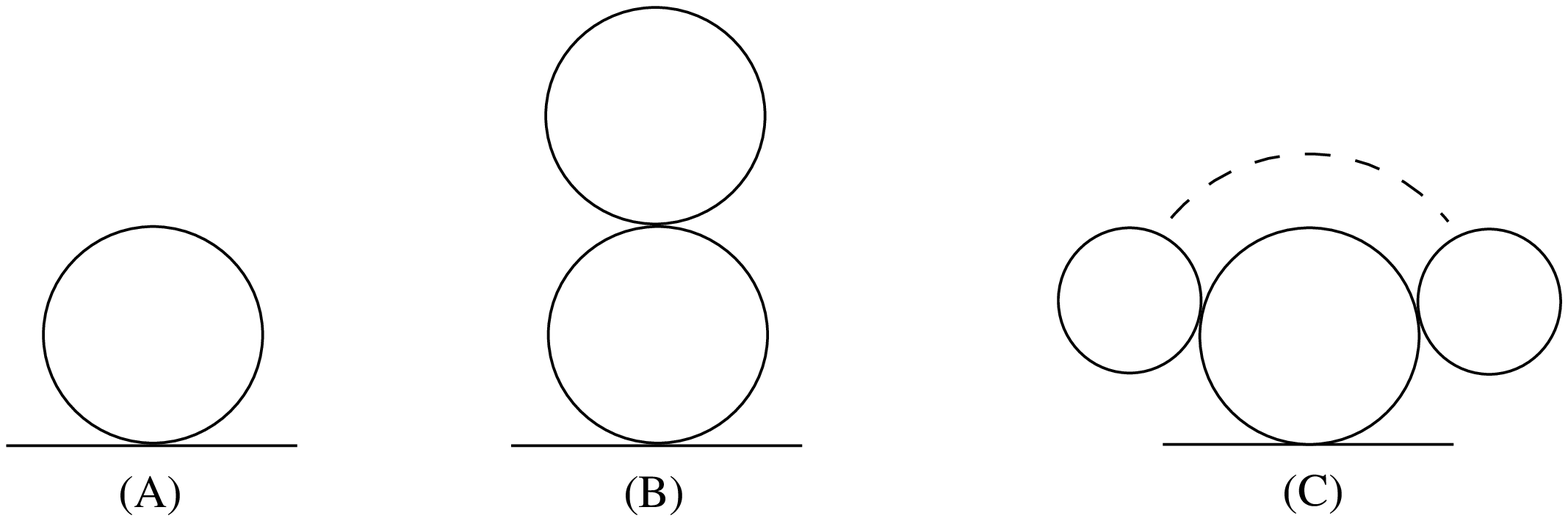}
}
     \caption{Bubble and cactus diagrams.}
 \label{tado}
\end{figure}

 Let us illustrate,  by using $\phi^4$ theory, the reason why 
 the naive perturbation theory at finite $T$ breaks down;
\beq
\label{toy}
  {\cal L}  =  {1 \over 2} [(\d \phi)^2 - \mu^2
  \phi^2] -{ \lambda \over 4! }\phi^4 .
\eeq
 We first consider the case $\mu^2 > 0$. The lowest order self-energy diagram
 Fig.\ref{tado} (A) is  $O( \lambda T^2)$ at high $T$. 
 However, Fig.\ref{tado} (B) is $O(\lambda T^2 \times {\lambda T \over \mu})$.
 Furthermore, higher powers of $T/\mu$ arise in higher 
 loops; e.g. the n-loop diagram in Fig.\ref{tado} (C) is 
 $O(\lambda^n T^{2n-1}/\mu^{2n-3})$.  Thus, the validity of the 
 perturbation theory breaks down when $T > \mu / \lambda$ because 
 the higher order diagrams are larger than lower ones. Therefore, 
 one should at least resum cactus diagrams to get sensible results 
 at high $T$  \cite{WEIN74}. Physics behind this resummation is 
 the well-known Debye screening mass in the hot plasma.  

 When $\mu^2 < 0$ and  the system has spontaneous symmetry breaking (SSB), 
 the naive perturbation shows another problem. 
 The tree-level mass $m_0$ in this case is defined as
\beq  
 m_0^2 =  \mu^2 + {\lambda \over 2} \xi^2(T) ,
\eeq
 where $\xi(T) $ is the thermal expectation value of $\phi$. 
 As $T$ increases, $\xi$ decreases. Then  $m_0^2$ 
 becomes negative (tachyonic) even before the critical temperature $T_c$
 is reached. If this happens, the naive perturbation using the tree-level
 propagator does not make sense and certain resummation
 should be carried out \cite{KL76}. Note that, for $T< T_c$, 
 there is no reason to believe that only the cactus diagrams 
 shown in Fig.1 are dominant; there exists a three-point 
 vertex $\lambda \xi  \phi^3$ which is not negligible for $T \sim \xi (T)$.

\subsection{Problems in self-consistent resummation methods}

 Self-consistent resummation method is a procedure
 to improve perturbation theory at finite $T$
 and to avoid the problems in Sec. \ref{N-RT}.
 However, the method has other difficulties \cite{renorm,okog}.

 In the naive perturbation theory, there arises no new UV divergences 
 at $T \neq 0$ because of the natural cutoff from 
 the Boltzmann distribution function. Therefore, all the 
 UV divergences at finite $T$ are canceled by the counter 
 terms prepared at $T=0$ \cite{prf}. 

 On the other hand, in self-consistent methods at $T \neq 0$,
 the situation is not that simple: In fact, 
 the tree-level propagators have $T$-dependent mass (such as 
 $m(T)$ in the above) which contains higher loop contributions
 through the self-consistent gap-equation \cite{renorm}.
 This leads to a necessity of $T$-dependent
 counter terms which are sometimes introduced in ad hoc ways.

 Another problem is the violation of the Nambu-Goldstone (NG) theorem:
 In many of the self-consistent methods, resummation with keeping 
 symmetry is a non-trivial issue, and the NG theorem is often violated.

\subsection{New resummation method}
\label{Newresum}

 For theories with SSB, loop-expansion rather than the 
 weak-coupling expansion is relevant, since one needs to treat 
 the thermal effective potential. Therefore, 
 we developed an improved loop-expansion at finite $T$ for 
 the purpose of resummation \cite{our}. 
 The method keeps the renormalizability and guarantees 
 the Nambe-Goldstone theorem order by order at finite $T$.

 In the following, we divide our resummation procedure into three steps
 and apply it to $\phi^4$ theory. The case for $O(N)$ $\phi^4$ theory 
 will be discussed in  Sec. \ref{N-theorem}.

 We start with the thermal effective action with an expansion
 parameter ``$\delta$'':
\beq
\label{naive-l}
  {\Gamma[\varphi^2]} = \ln  \int[d\phi]
  \exp \left[  {1 \over \delta} \int_0^{1/T} d^n x
  \left[ {\cal L} (\phi + \varphi;\mu^2) + J \phi
  \right] \right] ,	
\eeq
where $J \equiv - \d \Gamma[\varphi] /\d \varphi $ and
 $\int_0^{1/T} d^n x \equiv \int_0^{1/T} d\tau \int d^{n-1} x $.
 If we explicitly write $\hbar$ in eq.(\ref{naive-l}),
 it appears as ${1 \over \hbar} \int_0^{\hbar/T} d^4 x {\cal L} $.
 Therefore, the loop-expansion by $\delta$ at finite $T$
 does not coincide with the $\hbar$  expansion. 
 The expansion by $\delta$ should be regarded as a steepest 
 descent evaluation of the functional integral.

\vspace{0.2cm}

\centerline{\bf Step 1} 

 Start with a  renormalized Lagrangian  with counter terms
\beq
\label{ala1}
  {\cal L}(\phi;\mu^2) 
            & = & 
  {1 \over 2} [(\d \phi)^2 - \mu^2
  \phi^2] -{ \lambda \over 4! }\phi^4   \nonumber \\
& & +{1 \over 2}A(\d \phi)^2
    -{1 \over 2} B \mu^2 \phi^2  -{ \lambda \over 4! }C  \phi^4 
    + D \mu^4 .
\eeq
 Here we have explicitly written the argument $\mu^2$
 in ${\cal L}$ for later use. The $\overline{MS}$ scheme with 
 the dimensional regularization is assumed in (\ref{ala1}).
 Just for notational simplicity, the factor $\kappa^{(4-n)}$ to be
 multiplied to $\lambda$ is omitted ($\kappa$ is the renormalization point).

 The $c$-number counter term $D\mu^4$, which was not considered in \cite{BM},
 is necessary to make the thermal effective potential
 finite. Also, it plays a crucial role for
 renormalization in OPT as will be shown in Sec. \ref{R-OPT}.
 
 The renormalization constants are completely fixed at $T=0$ and 
 $\mu^2 > 0$, in which $A, B$, $C$ and $D$ are expanded as
\beq
\label{counterterms}
 \left(
 \begin{array}{c}  
 A \\ B \\ C \\ D
 \end{array} \right)
  = \sum_{l=1}^{\infty} 
 \left(
 \begin{array}{c}  
 a_l \\ b_l \\ c_l \\ d_l
 \end{array} \right)
 \delta^l.
\eeq
 Note that (i) the coefficients ($a_l, b_l, c_l, d_l$)
 are independent of $\mu^2$, since we use the mass independent 
 renormalization scheme, (ii)  the UV divergences in
 the symmetry broken phase $(\mu^2 < 0)$ can be removed by 
 the same counter terms determined for $\mu^2 > 0$ \cite{BW,kugo}, 
 and  (iii) $A, B, C$ and $D$ are independent of $T$ by definition.

 The relations of $A, B, C$ and $D$ with the standard 
 renormalization constants are $A = Z-1$, $B= Z_{\mu}Z-1$ 
 and $C = Z_{\lambda}Z^2 -1 $, where $Z$'s are defined by  
 $\phi_0 = \sqrt{Z} \phi$, $\lambda_0 =  Z_{\lambda} \lambda$ 
 and $\mu_0^2 = Z_{\mu} \mu^2$ with suffix $0$ indicating 
 unrenormalized quantities.

\vspace{0.2cm}

\centerline{\bf Step 2}

 Rewrite the Lagrangian (\ref{ala1})  by introducing
 a new mass parameter $m^2$ following the idea of OPT \cite{OPT}:  
\beq
\label{decomp}
 \mu^2 = m^2 - (m^2-\mu^2) \equiv m^2 - \chi. 
\eeq
 This identity should be used not only in the
 standard mass term but also in the counter terms \cite{point},
 which is crucial to show the order by order renormalization in OPT:
\beq
\label{toy2}
  {\cal L}(\phi; \mu^2)   
           & = & {\cal L}_{\rm r} + {\cal L}_{\rm c} \\
\label{toy20}
  {\cal L}_{\rm r}
           & = & {1 \over 2} [(\d \phi)^2 - m^2 \phi^2]
		-{ \lambda \over 4! }\phi^4 + {1 \over 2}\chi \phi^2 \\
\label{toy2c}
 {\cal L}_{\rm c} & = &
     {1 \over 2}A(\d \phi)^2
    -{1 \over 2} B (m^2 - \chi) \phi^2  -{ \lambda \over 4! }C  \phi^4 
   \nonumber \\
 & & + D (m^2-\chi)^2 .
\eeq
 $A$, $B$, $C$ and $D$ in  ${\cal L}_{\rm c}$ were already
 determined in Step 1.

 On the basis of eq.(\ref{toy2}), we define a ``modified'' 
 loop-expansion in which the tree-level propagator has a mass 
 $m^2 + \lambda \varphi^2 /2 $ instead of
 $\mu^2 + \lambda \varphi^2 /2 $.
 Major difference between this expansion and the 
 naive one is the following assignment,  
\beq
\label{assign}
 m^2 = O(\delta^0), \ \ \ \  \chi = O(\delta) .
\eeq
 The physical reason behind this assignment is the fact 
 that $\chi$ reflects the effect of interactions.
 If one makes an assignment, $ m^2 = O(\delta^0), \chi=O(\delta^0)$,
 the modified loop-expansion immediately reduces to the naive one.

 Since eq.(\ref{toy2}) is simply a reorganization of the Lagrangian, 
 any Green's functions (or its generating functional) calculated in the 
 modified loop-expansion should not depend on the arbitrary mass $m$
 if they are calculated in all orders.
 However, one needs to truncate  perturbation series at certain order
 in practice. This inevitably introduces explicit $m$ dependence in 
 Green's functions. Procedures to determine $m$ are given in Step 3 below. 
 \footnote{ One may generalize Step 2 by adding and subtracting
 $\alpha_0 (\d_0 \phi)^2$, $\alpha_1 (\d_i \phi)^2$ and $\gamma \phi^4$ 
 with $\alpha_0$, $\alpha_1$ and $\gamma$ being finite parameters 
 to be determined by the PMS or FAC conditions (see Step 3).
 The renormalizability can be also shown to be maintained in this case.
  However, 
 we will concentrate on the simplest version ($\alpha_{0,1} =\gamma =0$)
 in the following discussions.}

 To find the ground state of the system, one should look for 
 the stationary point of the thermal effective potential defined by
$ V(\varphi^2) 
  = - \Gamma[\varphi^2 = {\rm const.}] / \int_0^{1/T} d^4 x $.
 As mentioned above, $V$ calculated up to 
 $L$-th loops  $V_L(\varphi^2 ;m)$  has explicit $m$-dependence.
 Thus the stationary condition reads
\beq
\label{del}
  {\partial V_L(\varphi^2;m) \over \partial \varphi } =0 ,
\eeq
 where derivative with respect to  $\varphi$ does not act on $m$
 by definition. Eq.(\ref{del}) gives a stationary point 
 of $V_L$  for given $m$.

\vspace{0.2cm}

\centerline{\bf Step 3}

 The final step is to find an optimal value of $m$ by imposing physical 
 conditions  \`{a} la Stevenson \cite{PMS} such as the following.
\begin{itemize}
\item[(a)] The principle of minimal sensitivity (PMS): 
 this condition requires that a chosen quantity calculated up 
 to $L$-th loops  (${\cal O}_L$) should be stationary 
 by the variation of  $m$: 
\beq
\label{pmsc}
 {\partial {\cal O}_L \over \partial m } = 0 .
\eeq
 \item[(b)] The criterion of the  fastest apparent convergence (FAC): 
 this condition requires that the perturbative corrections in
 $O_L$ should be as small as possible for a suitable value of $m$.
\beq
\label{facc}
 {\cal O}_L - {\cal O}_{L-l} = 0,
\eeq
 where $l$ is chosen in the range, $1 \le l \le L $.
\end{itemize}

 The above conditions reduce to self-consistent gap equations whose 
 solution determine the optimal parameter $m$ for given $L$. 
 Thus $m$ becomes a non-trivial function of $\varphi$, $\lambda$ and 
 $T$. This together with the solution of (\ref{del}) completely
 determine the thermal expectation value $\xi(T) \equiv \langle \phi \rangle$ 
 as well as  the optimal parameter $m(T)$. Through this self-consistent 
 process, higher order terms in the naive loop expansion are resumed.

 The choice of ${\cal O}_L$ in Step 3 depends on the quantity one needs
 to improve most. To study the static nature of the phase transition, 
 the thermal effective potential $V_L(\varphi^2;m)$ is most relevant.
 Applying the PMS condition for $V_L$ reads
\beq 
\label{okopms}
  {\partial V_L(\varphi^2;m) \over \partial m } =0,
\eeq
 which gives a solution $m=m(\varphi)$. This can be used to
 improve the effective potential at finite $T$ as
 $ V_L(\varphi^2; m) \rightarrow  V_L(\varphi^2; m(\varphi))$.
 Also, $\xi(T)$ and $m(T)$ are obtained by solving (\ref{del}) 
 together with (\ref{okopms}). In this case, the following relation holds: 
 $  {d V(\varphi^2;m(\varphi))/ d \varphi }|_{\varphi = \xi} =
 {\d V(\varphi^2;m(\varphi)) / \d \varphi }|_{\varphi = \xi}$.

 To improve particle properties at finite $T$, it is more efficient 
 to apply PMS or FAC conditions directly to the two-point functions. 
 We will use FAC for the one-loop pion self-energy in Section III 
 to show its usefulness.

\subsection{Renormalization in  OPT }
\label{R-OPT}

 We now prove the order by order renormalization in OPT. 
 Let us first rewrite eq.(\ref{toy2}) as
\beq
\label{ala2}
  {\cal L}(\phi;\mu^2) & = & {\cal L}(\phi;m^2)  \nonumber \\
  &  &  + {1 \over 2} \chi \phi^2 +\left[ {1 \over 2} B\chi \phi^2
  + D \chi^2 - 2D m^2 \chi \right] .
\eeq
 The UV divergences arising in the perturbation theory are classified 
 into two classes: The divergences in the Green's function generated 
 by ${\cal L}(\phi; m^2)$, and the divergences obtained by the multiple 
 insertion of  $(1/2) \chi \phi^2$ to the Green's function generated 
 by  ${\cal L}(\phi; m^2)$.

 Since we use the symmetric and mass independent renormalization scheme 
 (such as the $\overline{MS}$ scheme), any divergences in the first class are
 renormalized solely by the coefficients $A$, $B$, $C$ and $D$
 in ${\cal L}(\phi; \mu^2)$.
 Although $T$-dependent divergences appear because of the $T$-dependent
 ``resumed'' mass $m^2(T)$, they are properly renormalized away 
 since the counter terms (such as $Bm^2\phi^2$,  $Dm^2$ and $Dm^4$) 
 also acquires $T$-dependence through $m^2(T)$. 
 In other words, the divergences arising from the resumed
 propagator is removed by the resumed  counter terms. (See also,
footnote 1.)

\begin{figure}[h]
\centerline{
    \epsfxsize=7.4cm
    \epsfbox{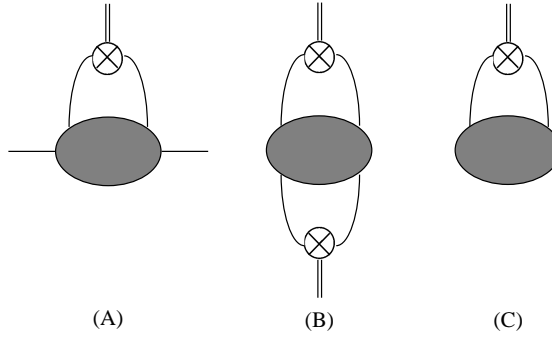}
}
     \caption{Diagrams which contain UV divergences as a result
     of the multiple insertion of $(1/2)\chi \phi^2$.
    (A) corresponds to a single insertion with two external lines.
    (B) and (C) have no external lines with
     a single insertion and a double insertion, respectively.}
 \label{comr}
\end{figure}

 The divergences in the second class can be  shown to be removed 
 by  the last three counter terms in (\ref{ala2}).
 (Note that $B$ and $D$  are already fixed in Step 1, and we do not 
 have any freedom to change them.) This is obviously related to the 
 renormalization of composite operators. In fact, the standard 
 method \cite{IZ} tells us that necessary counter terms are written as 
\beq
\label{count2}
 {1 \over 2 } (Z Z^{-1}_{\phi^2} -1) \chi \phi^2 + \Delta_2 \chi^2 
  + \Delta_1 \chi . 
\eeq
 Here $Z_{\phi^2}$ is the renormalization constant for the
 composite operator $\phi^2$, and removes the divergence 
 in Fig.\ref{comr}(A). $\Delta_2$ and $\Delta_1$ are necessary 
 to remove the overall divergences in Fig.\ref{comr}(B) 
 and in Fig.\ref{comr}(C), respectively. 

 Now, one can prove that (\ref{count2}) coincides with 
 the last three terms in (\ref{ala2}):
\beq
\label{eqq}
  Z Z^{-1}_{\phi^2} -1 = B, \ \ \ \Delta_2 = D, \ \ \ {\rm and } \ \ \ 
  \Delta_1 = -2Dm^2.
\eeq
 The first equation is obtained by the definition $B= Z_{\mu}Z-1$ 
 and an identity 
\beq
\label{zet1}
 Z_{\phi^2} = Z_{\mu}^{-1}.
\eeq
 The overall divergence of the vacuum diagram with
 no external-legs is removed by the $c$-number counter
 term $Dm^4$ in ${\cal L}(\phi; m^2)$. Therefore,  
 the last two equations in (\ref{eqq}) are obtained as
\beq
\label{zet2}
 \Delta_1 & =  &
  - ({\partial \over \partial m^2}) \left[ Dm^4 \right] = -2D m^2 ,\\ 
\label{zet3}
 2 \Delta_2 & = &
  ({\partial \over \partial m^2})^2 \left[ Dm^4 \right] = 2D.
\eeq
 Eq.(\ref{eqq}) shows clearly that all the necessary counter terms in OPT
 are supplied solely by the original Lagrangian ${\cal L}(\phi; \mu^2)$.  
 Thus, we can carry out renormalization order by
 order even within the self-consistent method. For more detailed
 proof of the relations (\ref{eqq}), see Appendix A of ref.\cite{our}. 

\vspace{0.2cm}

Three comments are in order here:
\begin{itemize}
\item[(i)] Because the renormalization is already carried out
 in Step 2, one obtains finite gap-equations from the beginning
 in Step 3.  Our procedure ``resummation after renormalization''
 has many advantages over the conventional procedure 
 ``resummation before renormalization'' where UV divergences 
 are hoped to be canceled after imposing the gap-equation.
 The difference between the two is prominent 
 especially in higher order calculations. 
\item[(ii)]
 The decomposition (\ref{decomp}) should be done both
 in the mass term and the counter terms. This guarantees order by order
 renormalization in our modified loop-expansion in any higher orders.
 (In ref.\cite{BM}, the renormalizability was checked up to 
 the two-loop level in the $\phi^4$ theory at high $T$.)
 On the other hand, if one keeps the original counter 
 term $(1/2) B \mu^2 \phi^2 + D \mu^4$ without the decomposition 
 (\ref{decomp}), $L$-loop diagrams with $L > M$ must be taken into account 
 to remove the UV divergences in the $M$-loop order.
 This is an unnecessary complication due to the inappropriate 
 treatment of the counter terms. 
 (See e.g. ref.\cite{KPP} which encounters this  problem). 
\item[(iii)]
 As far as we stay in the low energy region far below the
 Landau pole, we need not address the issue of the triviality of 
 the $\phi^4$ theory \cite{triv}: Perturbative renormalization in OPT works
 in the same sense as that in the naive perturbation.

\end{itemize}

\subsection{Nambu-Goldstone theorem in OPT}
\label{N-theorem}

 The procedure and the renormalization in OPT  discussed above
 do not receive modifications even if the Lagrangian has global symmetry.
 For $O(N)$ $\phi^4$ theory, one needs to replace $\phi$ and $\phi^2$ 
 by $\vec{\phi} = (\phi_0, \phi_1, \cdot \cdot \cdot , \phi_{N-1})$
 and $\vec{\phi}^2$ respectively in all the previous formulas.

 In the symmetry broken phase of such theory,
 the Nambu-Goldstone (NG) theorem and massless
 NG bosons are guaranteed in each order of the modified 
 loop-expansion in OPT for arbitrary $N$. 
 To show this, it is most convenient to start
 with the thermal effective potential $V(\vec{\varphi}^2)$.
 By the definition of the effective potential, $V(\vec{\varphi}^2)$ has
 manifest $O(N)$ invariance if it is calculated in all orders.
 
 In OPT, $V$ calculated up to $L$-th loops $V_L(\vec{\varphi}^2;m)$
 has also manifest $O(N)$ invariance, because our 
 decomposition (\ref{decomp}) used in  (\ref{toy2}) does not 
 break $O(N)$ invariance. Once $V_L$ has invariance under the $O(N)$ rotation
 ($\varphi_i \rightarrow \varphi_i + i \theta^a T^a_{ij} \varphi_j$),
 the immediate consequence is the standard identity: 
\beq
\label{NG1}
 {\d V_L(\vec{\varphi}^2;m) \over \d \varphi_j } T_{ji}^a
 = - {\d^2 V_L(\vec{\varphi}^2;m) \over \d \varphi_i \d \varphi_j } 
 T_{jk}^a \varphi_k,
\eeq
 with ${\bf T}^a $ being the generator of the $O(N)$ symmetry.
 Eq.(\ref{NG1})  is valid for arbitrary $L$, $m$ and $N$.

 At the stationary point where the l.h.s. of (\ref{NG1})
 vanishes, there arises  massless NG bosons
 for $T_{jk}^a \varphi_k \neq 0$, since the r.h.s. of (\ref{NG1}) 
 is equal to $ -{\cal D}_{ij}^{-1}(0) T_{jk}^a \phi_k $ where
 ${\cal D}_{ij}(0)$ is the Matsubara propagator at zero frequency 
 and momentum calculated up to $L$-th loops. Thus the existence of the
 NG bosons is proved independent of the structure of the gap-equation 
 in Step 3.

 Now, let us show an example of the unjustified approximations leading to 
 the breakdown of the NG theorem. 
 Suppose that we make a general decomposition such as 
\beq
\label{abs}
  \mu^2 \delta_{ij} = m^2_{ij} - ( m^2_{ij} - \mu^2 \delta_{ij}),
\eeq
 with $m^2_{ij} \neq m^2 \delta_{ij}$. This leads to an $O(N)$ 
 non-invariant effective potential, and the relation (\ref{NG1}) 
 is not guaranteed in any finite orders of the loop-expansion.
 For example, when the $O(N)$ symmetry is spontaneously broken 
 down to $O(N-1)$, one may be tempted to make a decomposition
 $m_{ij}^2 = m_0^2 $ ($i=j=0$), $m_{ij}^2 = m_1^2 $ ($i = j \neq 0$),
 and $m_{ij}^2 = 0 $ ($i \neq j$) to impose self-consistent conditions
 for the radial mode and the NG mode. However, the effective potential 
 does not have $O(N)$ symmetry in this case and eq.(\ref{NG1}) does not hold. 

\section{Application to $O(4)$ $\sigma$ model}

 Let us apply OPT to the $O(4)$ $ \sigma$ model.
 The model shares common symmetry and dynamics with QCD and 
 has been used to study the real-time dynamics and critical phenomena 
 associated with the QCD chiral transition \cite{CH,PW}.

\subsection{Parameters at $T=0$}

 The $O(4)$ $\sigma$ model reads
\beq
\label{lin1}
  {\cal L} & = & 
  {1 \over 2} [(\d \vec{\phi})^2 - \mu^2 \vec{\phi}^2]
  -{ \lambda \over 4! }(\vec{\phi}^2)^2 + h \sigma  \nonumber \\
& &\mbox{}+{1 \over 2} A(\d \vec{\phi})^2 -{1 \over 2} B \mu^2 \vec{\phi}^2 
  -{ \lambda \over 4! }C(\vec{\phi}^2)^2+D \mu^4 ,
\eeq
 with $\vec{\phi}=(\sigma, \vpi)$.
 $h \sigma$ is an explicit symmetry breaking term.

 $A$, $B$, $C$ and $D$ in the one-loop order are
\beq
\label{counter}
  A =0, \ \ \ 
  B = {\lambda \over 16\pi^2} \bep , 
  \ \ \ C={\lambda \over 8\pi^2}\bep , 
  \ \ \  D = -{1 \over 16 \pi^2} \bep ,
\eeq
 where $ \bep \equiv {2 \over 4-n}-\gamma+\log(4\pi)$,
 with $\gamma$ being the Euler constant.

 When SSB takes place $(\mu^2 < 0)$, the 
 replacement $\sigma \rightarrow \sigma + \xi$ in  eq.(\ref{lin1}) 
 leads to the tree-level masses of $\sigma $ and $\pi$;
\beq
\ms ^2= \mu^2+\frac{\lambda}{2}\xi^2, \ \ 
\mp ^2= \mu^2+\frac{\lambda}{6}\xi^2 .
\eeq
 The expectation value $\xi$ at $T=0$ is determined by the stationary 
 condition for the standard effective potential 
 $\d V(\vec{\varphi})/\d \varphi_j =0 $.

 Later we will take a special FAC condition in which 
 $m^2$ deviates from $ \mu^2$ only at $T \neq 0$, so that
 the naive loop-expansion  at $T=0$ is valid.
 The renormalized couplings $\mu^2, \lambda$ and $h$
 can thus be determined by the renormalization conditions 
 in the naive loop-expansion at zero $T$ such as 
 (i) $m_{\pi}= 140$ MeV,
 (ii) $f_{\pi} = 93 $ MeV, and
 (iii)  $\pi$-$\pi$ scattering phase shift  \cite{CH}.

 Instead of (iii), one may adopt $m^{peak}_{\sigma}$ 
 (the peak position of the spectral function in the $\sigma$ channel). 
 We take this simplified condition with three possible cases: 
 $m^{peak}_{\sigma}$ =550 MeV, 750 MeV and 1000 MeV.
 $m^{peak}_{\sigma}=$ 550 MeV and 750 MeV are consistent with 
 recent re-analyses of the $\pi$-$\pi$ scattering phase shift \cite{pipi}. 

\vspace{0.2cm}

\begin{table}
\[ 
\begin{array}{|c||c|c|c|c|} \hline 
  \ m_{\sigma}^{peak} \ (\mev) 
& \mu^2 \ (\mev^2) \ & \  \lambda  & h \ (\mev^3) \ &
  \  \kappa \ (\mev) \ \\ \hline
  550 & -284^2 & 73.0 & 123^3 & 255 \\
  750 & -375^2 & 122  & 124^3 & 325 \\
 1000 & -469^2 & 194  & 125^3 & 401 \\ \hline
\end{array} 
\]
\caption{Vacuum parameters corresponding to $m_{ \sigma}^{peak}
 =$ 550, 750, 1000 MeV} 
\label{tab1}
\end{table}

 We still have a freedom to choose the renormalization point $\kappa$.
 Instead of trying to determine optimal $\kappa$ by the renormalization 
 group equation for the effective potential \cite{kast}, 
 we take a simple and physical condition $m_{0\pi}$=$m_{\pi}$=140 MeV. 
 This choice has two advantages: (a) One-loop pion 
 self-energy $\Sigma_{\pi}(k^2)$ vanishes at the tree-mass; 
 $\Sigma_{\pi}(k^2 = m_{0 \pi}^2) = \Sigma_{\pi}(k^2=m_{\pi}^2)=0$, and
 (b) the spectral function in the $\sigma$ channel
 starts from a correct continuum threshold in the one-loop level. 
 Resultant  parameters are summarized  in TABLE I.

\begin{figure}[h]
\centerline{
    \epsfysize=6.3cm
    \epsfbox{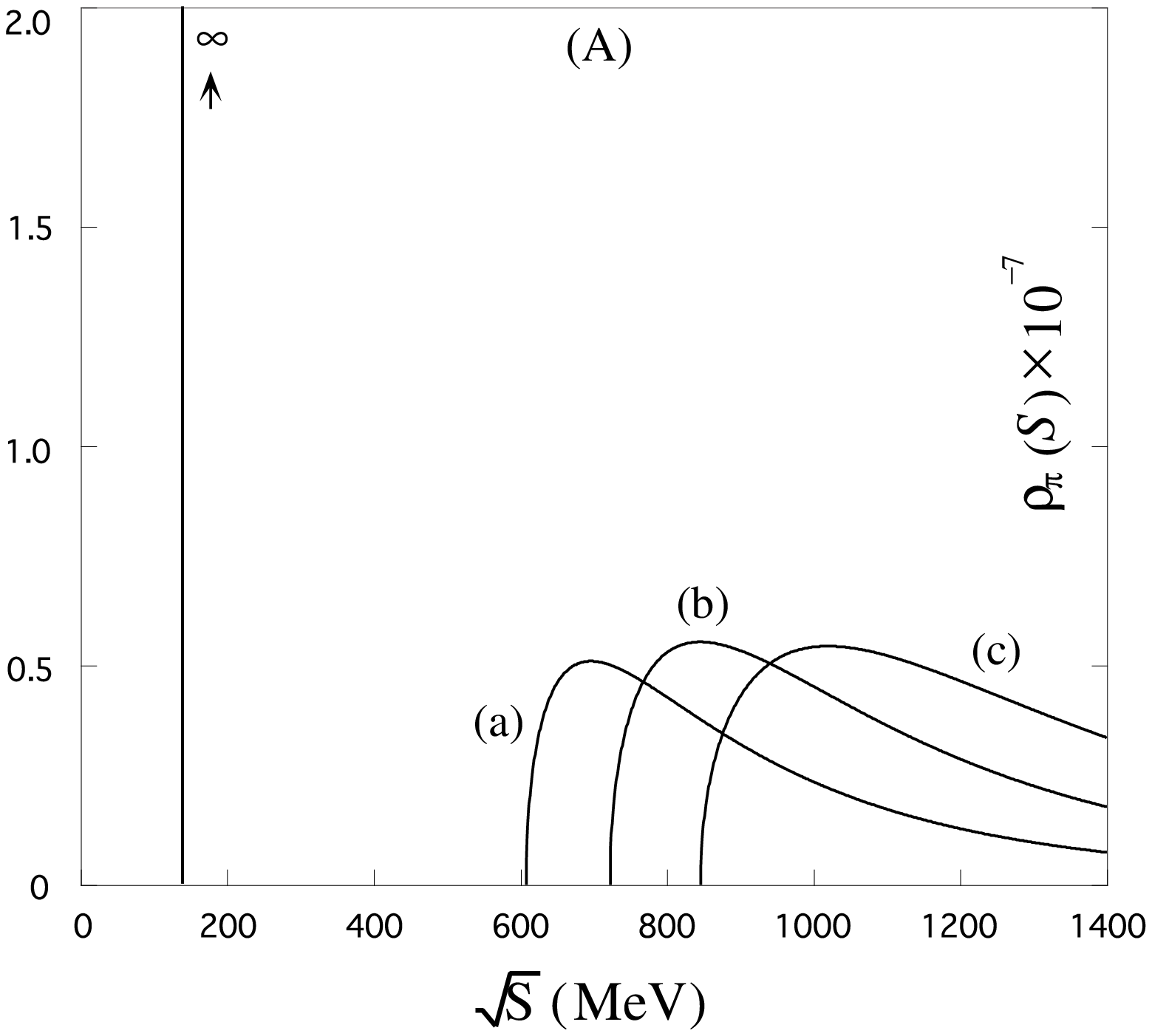} \hspace{0.2cm}
    \epsfysize=6.3cm
    \epsfbox{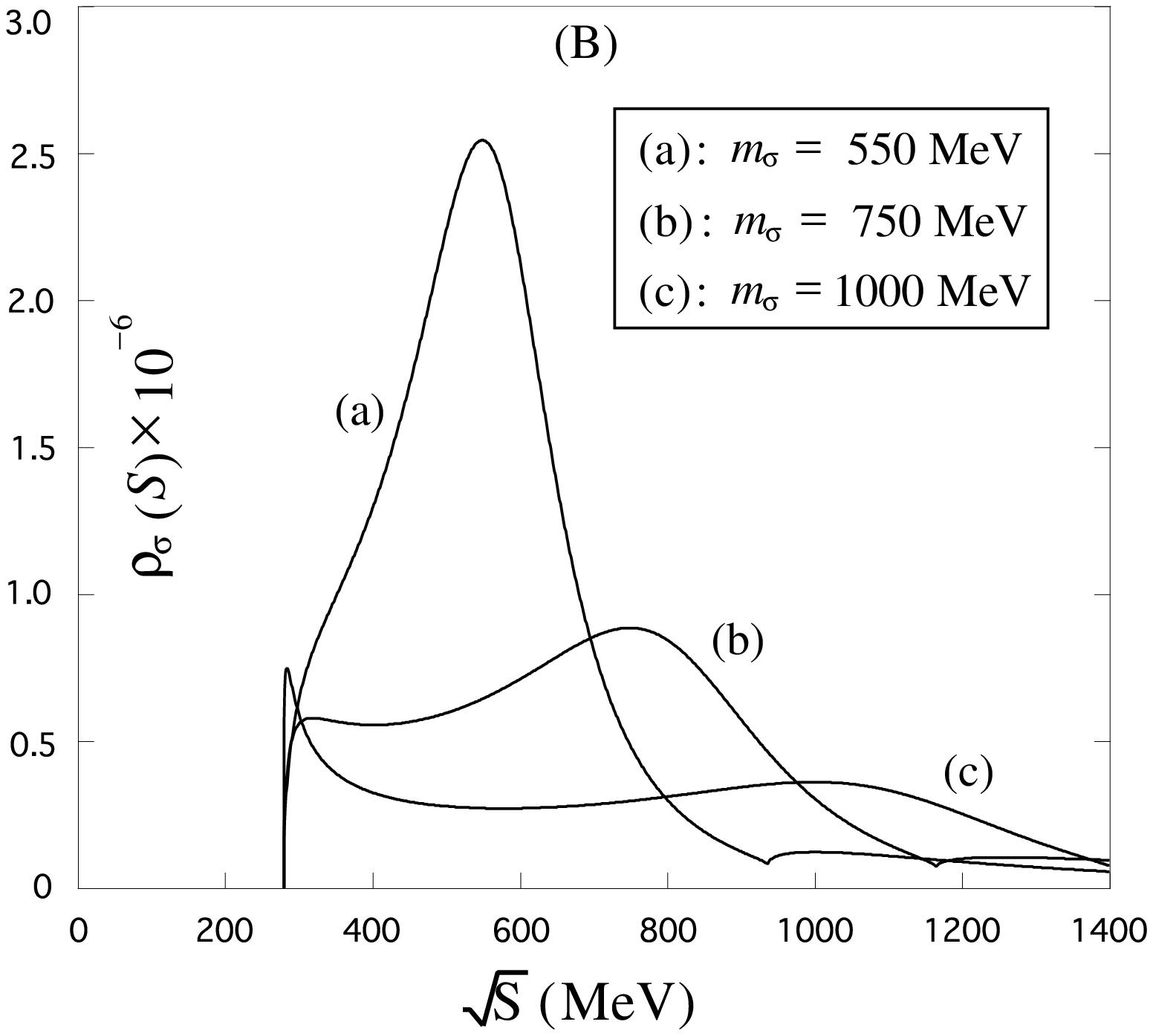}
}
     \caption{Spectral functions at $T=0$ in the 
     $\pi$ channel (A) and in the $\sigma$ channel
     (B) for $m^{peak}_{\sigma}=550$ MeV, 750 MeV and  1000 MeV.}
 \label{spect}
\end{figure}

 In Fig.\ref{spect}, the spectral functions $\rho_{\sigma}$ 
 and $\rho_{\pi}$ at $T=0$, namely the $T=0$ limit of 
 eq.(\ref{spectral}) defined below, are shown as a function of
 $\sqrt{s} \equiv \sqrt{\omega^2 - {\bf k}^2}$.
 In the $\pi$ channel, there are one particle pole and 
 a continuum  starting from the threshold $\sqrt{s_{th}}
 = m_{0 \pi} + m_{0 \sigma}$. $\sqrt{s_{th}}$ 
 is the point where the channel $\pi + \sigma$ opens.
 In the $\sigma$ channel, the spectral function starts
 from the threshold $2 m_{0 \pi} = 280$ MeV and shows 
 a broad peak centered around $\sqrt{s} = m^{peak}_{\sigma}$.
 The large width of $\sigma$ is due to a strong $\sigma$-$2\pi$ coupling  
 in the linear $\sigma$ model. The corresponding $\sigma$-pole
 is located far from the real axis on the complex $s$ plane.

 Here we show the definition of the spectral function at finite $T$:
\beq
\label{spectral}
   \rho_{\phi}(\omega, {\bf k};T)  = 
   - {1 \over \pi} \mbox{Im} D_{\phi}^{R}(\omega, {\bf k};T) ,
\eeq
where $D_{\phi}^{R}$ is the retarded correlation function
\beq
  D_{\phi}^{R}(\omega, {\bf k};T) 
    =  -i  \int d^4x e^{i kx}
   \theta(t) \la  [\phi(t,{\bf x}), \phi(0,{\bf 0})] \ra,
\eeq  
 with  $\la \cdot \ra$ being the thermal expectation value.

\subsection{Application of OPT}

 Now let us proceed to Step 2 in OPT and rewrite eq.(\ref{lin1}) as 
\beq
\label{lag2}
  {\cal L} & = & {1 \over 2} [(\d \vec{\phi} )^2 -m^2 \vec{\phi}^2]
   -{ \lambda \over 4! } (\vec{\phi}^2)^2 
   + {1 \over 2} \chi \vec{\phi}^2 
  + h \sigma \nonumber\\
  & &  - {1 \over 2} B m^2 \vec{\phi} ^2 
 -{ \lambda \over 4! } C (\vec{\phi}^2)^2 
       + D m^4.
\eeq
 Since $\chi$ ( = $m^2-\mu^2$) is already a one-loop order,
 we have neglected the terms proportional to $B \chi$, $D \chi^2$ 
 and $D \chi$ which are two-loop or higher orders.

 When SSB takes place ($\vec{\phi} \rightarrow \vec{\phi} + \vec{\varphi}$),
 the tree-level masses to be used in the modified loop-expansion read
\beq
\label{Rbmass}
  \ms ^2=m^2+\frac{\lambda}{2}\vec{\varphi}, \ \  
  \mp ^2=m^2+\frac{\lambda}{6}\vec{\varphi}.
\eeq
 Since $m^2$ will eventually be a function of $T$, the tree-masses 
 running in the loops are not necessary tachyonic at finite $T$ 
 contrary to the naive loop-expansion (see the discussion in Sec. \ref{N-RT}).

 The thermal effective potential $V(\vec{\varphi};m)$ is calculated 
 in the standard manner except for the extra terms proportional to $\chi$.
 The effective potential in the one-loop level reads
\beq
\label{pot}
   V(\vec{\varphi};m) & = & {1 \over 2} \mu^2 \vec{\varphi}^2 + 
   {\lambda \over 4!}\vec{\varphi}^4 - h \vec{\varphi} \nonumber \\
 & + & {1 \over 64 \pi^2} \left[ m_{0 \sigma}^4 \  
 {\rm ln} \left| { m_{0 \sigma}^2 \over \kappa^2 e^{3/2} } \right|
  + 3 m_{0 \pi}^4 \ 
  {\rm ln} \left| { m_{0 \pi}^2 \over \kappa^2 e^{3/2}} \right|   
 \right]
 \\
 & + & T \int {d^3k \over (2 \pi)^3} \left[ {\rm ln} 
  (1- e^{- E_{\sigma} /T} ) + 3\  {\rm ln} 
  (1- e^{- E_{\pi} /T} ) \right] \nonumber,
\eeq
 where $E_{\phi} \equiv \sqrt{ {\bf k}^2 + m_{0 \phi}^2 }$.
 Although this has the similar structure with the standard free energy
 in the naive loop-expansion, the coefficient of the first term 
 in the r.h.s. of (\ref{pot}) is $\mu^2$ instead of $m^2$. 
 This is because we have extra mass-term proportional to
 $\chi$ in the one-loop level.  The stationary point $\xi$  
 is obtained by 
\beq
\label{dvev}
 \left. {\d V(\vec{\varphi} ;m) \over \d \varphi_i} \right| _{\vec{\varphi} 
 = (\xi, {\bf 0})} =0.
\eeq
 Since the derivative with respect to $\xi$ does not act on $m$,
 this gives a solution $\xi$ as a function of $T$ and $m$.
 By imposing another condition on $m$ (Step 3), one eventually
 determines both $\xi$ and $m$ for given $T$.

\subsection{FAC condition for $m^2$}

 To resum the hard thermal loops, the PMS condition for the effective 
 potential requires 2-loop calculation, while the FAC condition for 
 the self-energy requires only 1-loop calculation. Therefore, we adopt 
 the latter condition here  to determine $m^2$.

 The retarded self-energy $\Sigma^R_{\phi}$ (defined by
 $[D_{\phi}^R]^{-1} = s - m_{0\phi}^2 - \Sigma^R_{\phi}$) 
 is related to the 11-component of the 2 $\times$ 2 self-energy 
 in the real-time formalism \cite{real};
\beq
  \label{R-11}
  {\rm Re} \Sigma^R_{\phi} (\omega, {\bf k};T)& = &
  {\rm Re} \{\Sigma^{11}_{\phi}(\omega,{\bf k})+
  \Sigma^{11}_{\phi}(\omega, {\bf k};T)\}  \\ 
  {\rm Im} \Sigma^R_{\phi} (\omega, {\bf k};T)& = &
  \tanh({\omega \over 2T})\  {\rm Im} \{\Sigma^{11}_{\phi}(\omega,{\bf k})+
  \Sigma^{11}_{\phi} (\omega, {\bf k};T)\}. \nonumber
\eeq
 Here  $\Sigma^{11}_{\phi}(\omega,{\bf k};T)$ is a part having 
 explicit $T$-dependence through the Bose-Einstein distribution, 
 while $\Sigma^{11}_{\phi} (\omega, {\bf k})$ is a part having 
 only implicit $T$-dependence through $m(T)$ and $\xi(T)$. 
 In the one-loop level, $\Sigma^{11}_{\phi}$ can be calculated 
 only by the 11-component of the free propagator,
\beq
\label{T-pro}
  iD_{0 \phi}^{11}(k^2;T) =  {i \over k^2-m_{0 \phi}^2+i\epsilon} 
   + 2 \pi n_{B} \delta(k^2-m_{0 \phi}^2),
\eeq
 with $n_{B}=[e^{\omega /T}-1]^{-1}$.

\begin{figure}[h]
\centerline{
    \epsfxsize=18.0cm
     \epsfbox{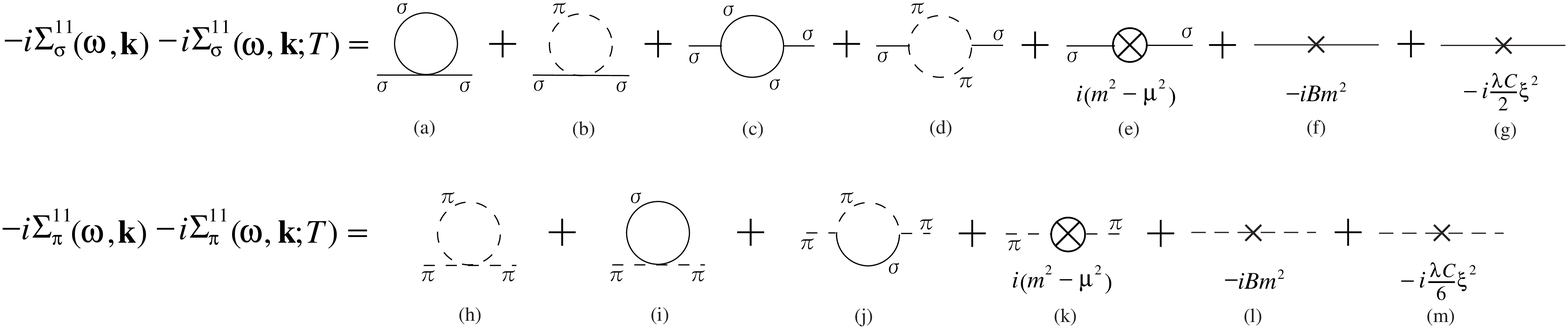}
}
     \caption{One-loop self-energy  $\Sigma^{11}$ for $\sigma$ and 
     $\pi$ in the  modified loop-expansion at finite $T$.}
 \label{self}
\end{figure}

 One-loop diagrams in OPT for $\Sigma^{11}_{\phi}$ are shown in 
 Fig. \ref{self}. Their explicit forms are given in \cite{our}.
 The NG theorem discussed in Sec. \ref{N-theorem} can be explicitly checked
 by comparing eq.(\ref{dvev}) and the inverse pion-propagator 
 at zero momentum $[D^{R}_{\pi}(0,{\bf 0};T)]^{-1} $.

 Let us impose the FAC condition on $\Sigma_{\pi}^R$.
 Since we chose a renormalization condition $m_{0\pi} = m_{\pi} =140$ MeV
 at zero $T$, one may be tempted to adopt the following condition at finite
 $T$:
\beq
\label{kap}
  \Sigma_{\pi}^R ( \omega=m_{0 \pi}, {\bf 0}; T) = 0.
\eeq
 However, eq.(\ref{kap}) does not guarantees that $m^2$ is real,
 since the l.h.s. of eq.(\ref{kap}) receives an imaginary part due to the 
 Landau damping. To avoid this problem, we take a hybrid condition:
\beq
\label{CHC}
 \Sigma_{\pi}^{11} ( \omega=m_{0 \pi}, {\bf 0}) 
 +  \Sigma_{\pi}^{11} ( \omega=0, {\bf 0}; T) = 0.
\eeq
Note that the external energy is set to be zero in  the $T$-dependent
part.\footnote{Eq. (\ref{CHC}) can be formulated in a covariant way 
 by introducing the four-vector $n_{\mu}$ characterizing the heat bath.
 The first term of the equation  is  only a function of $k_{\mu}^2$ 
 because it is $T$ independent. The second term is a function of
 $k_{\mu}^2$ and  $k_{\mu} n^{\mu}$. 
 Therefore, the condition reads
  $ \Sigma_{\pi}^{11} ( k_{\mu}^2 = m_{0 \pi}^2) 
 +  \Sigma_{\pi}^{11} ( k_{\mu}^2 = 0,  k_{\mu} n^{\mu} =0; T) = 0.$}
  Because the second term in the l.h.s. vanishes
 at $T=0$, the solution  of eq.(\ref{CHC}) at $T=0$ becomes 
\beq
\label{self1}
  m^2(T=0) = \mu^2.
\eeq 
 Therefore, the FAC condition (\ref{CHC}) does not spoil the naive-loop 
 expansion at $T=0$. 

 At high $T$ ($\xi(T) \simeq 0$), the following analytic solution 
 is obtained as far as $T^2 \gg m^2$;
\beq
\label{hightt}
  m^2 (T) = \mu^2 + {\lambda \over 12} T^2 ,
\eeq
 which implies that the Debye screening mass at high $T$ is 
 properly taken into account. For realistic values of $\lambda$ 
 in TABLE I, the condition $T^2 \gg m^2$ is not well satisfied and 
 eq.(\ref{CHC}) should be solved numerically.
 
 For intermediate values of $T$, eq.(\ref{CHC}) can effectively
 sum not only the contributions from the diagrams 
 in Fig.\ref{self}$(a,b,h,i)$, but also from those in Fig.\ref{self}$(c,d,j)$.
 Thus, OPT can go beyond the cactus approximation which sums 
 only the diagrams in Fig.\ref{self}$(a,b,h,i)$.

\vspace{0.2cm}

 Two remarks are in order here.

\begin{enumerate}
\item[(i)] By eq.(\ref{CHC}), only the ${\bf k}$-independent part
 of the self-energy is resumed. If one needs to resum
  ${\bf k}$-dependent part too, one must introduce  $m^2$ which 
 depends both on $T$ and ${\bf k}$ and impose ${\bf k}$-dependent 
 FAC or PMS conditions\cite{confref}.
\item[(ii)] 
 In ref.\cite{KPP}, it has been studied the convergence properties of 
 the free energy at high $T$ with a variational condition equivalent
 to the PMS condition here.
 Although the approach has a problem of renormalization as we have 
 already mentioned, the result is suggestive in the sense that
 the optimized loop expansion has much better convergence properties than 
 the loop expansion based on the hard thermal loops \cite{confref2}. 
 Better understanding of the convergence properties
 both in PMS and FAC conditions is an important future problem.
\end{enumerate}

\subsection{Behavior of $m(T)$, $m_{0 \phi}(T)$ and $\xi(T)$}
\label{result}

\begin{figure}[h]
\centerline{
    \epsfysize=6.3cm
    \epsfbox{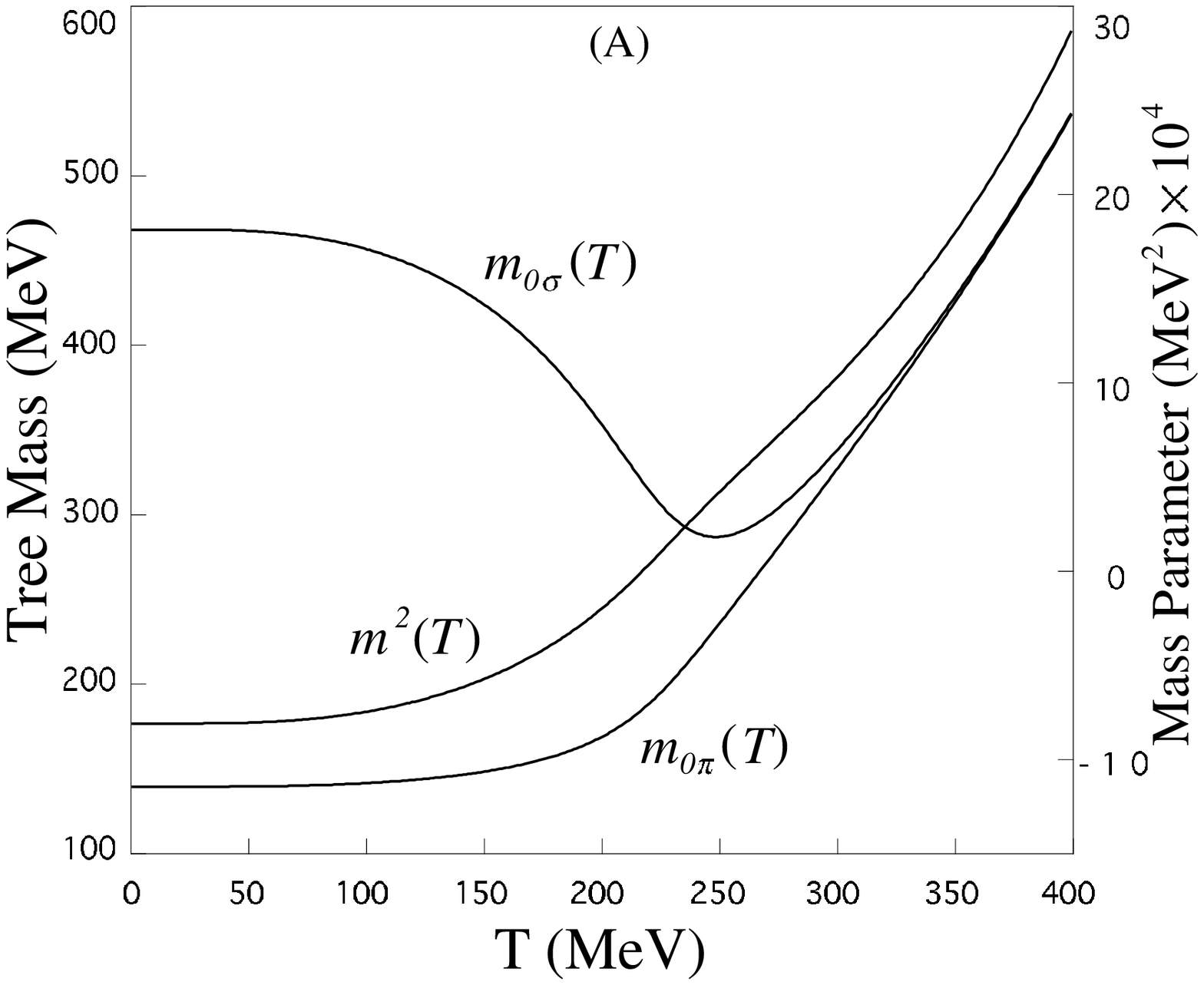}\hspace{0.2cm}
    \epsfysize=6.3cm
    \epsfbox{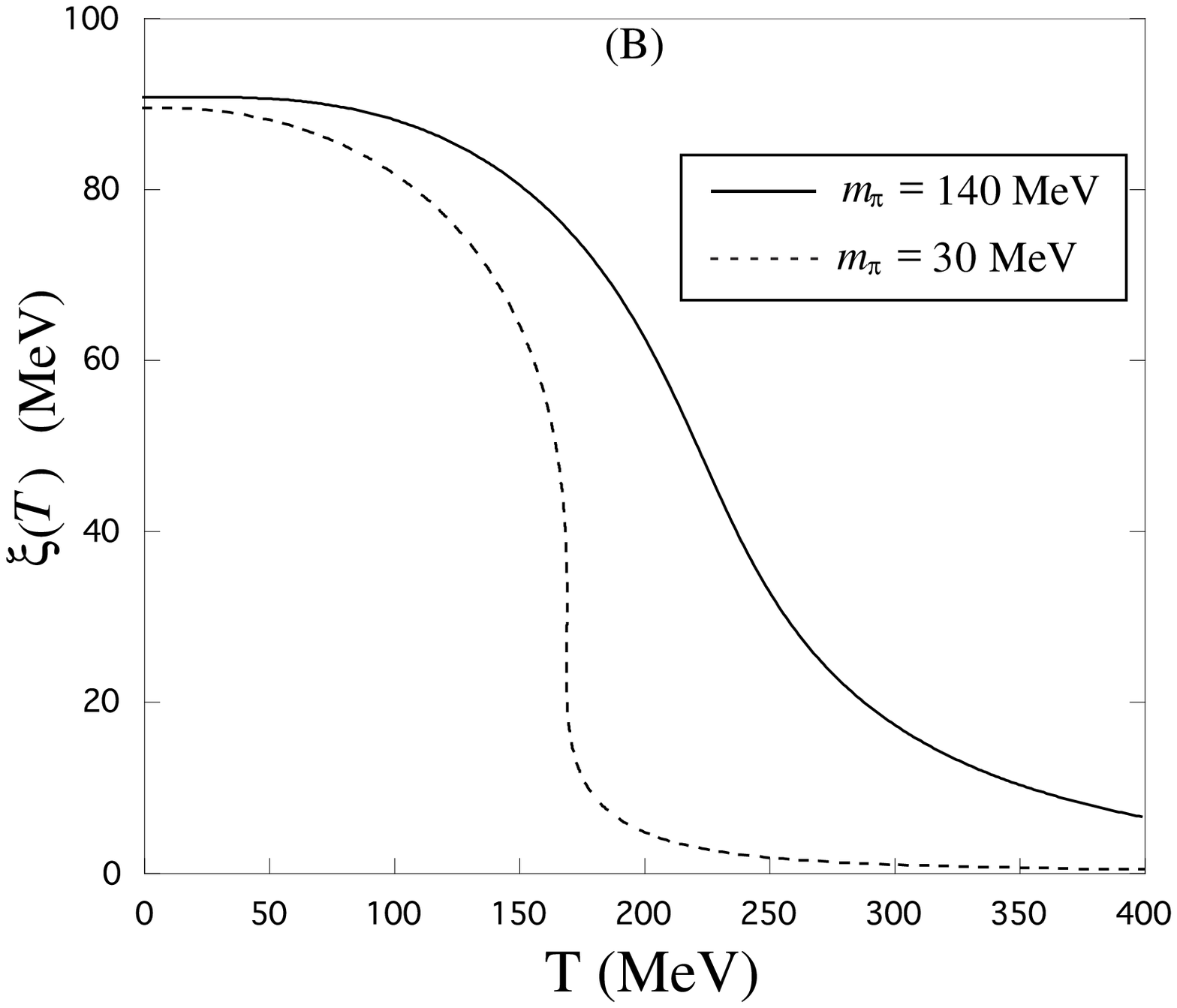}
}
     \caption{(A) Masses in the tree-level $\mp(T)$ and 
     $\ms(T)$ shown with left vertical scale, and 
     the mass parameter $m^2(T)$ with the right vertical scale.
     (B) $\xi(T)$ for $m_{\pi}(T=0)=140 $MeV and 30 MeV with 
     $m^{peak}_{\sigma}(T=0)=550$ MeV.}
 \label{vevm}
\end{figure}

 In Fig.\ref{vevm}(A) the tree-level masses in eq.(\ref{Rbmass})
 and $m^2(T)$ are shown for $m_{\sigma}^{peak} (T=0)=550$ MeV. 
 $m_{0\phi}^2(T)$ is not tachyonic and approaches to $m^2(T)$ 
 in the symmetric phase. This confirms that our resummation procedure 
 cures the problem of tachyons  in Sec. \ref{N-RT}.

 The solid line in Fig.\ref{vevm}(B) shows the chiral 
 condensate $\xi(T)$ obtained by minimizing the effective potential
 with  $m_{\sigma}^{peak}(T=0) = 550$ MeV.
 $\xi(T)$ decreases uniformly as $T$ increases, which
 is a typical behavior of the chiral order parameter
 at finite $T$ away from the chiral limit.

 As we approach the chiral limit ($h \rightarrow 0$ or
 equivalently $m_{\pi} \rightarrow 0$), 
 $\xi(T)$ develops multiple solutions for given
 $T$, which could be an indication of the first order transition.  
 The critical value of the quark mass $m_q^{\rm crit.}$
 below which the multiple solutions arise is
\beq
   m_q^{\rm  crit.} / m_q^{\rm phys.} = 
 (m_{\pi}^{\rm crit.}  / m_{\pi}^{\rm phys.})^2 = 0.08,
\eeq
 where we have used Gell-Mann-Oakes-Renner relation \cite{GOR}
 to related the pion mass with the quark mass.
 $m_q^{\rm phys.}$ is the physical light-quark mass 
 corresponding to $m_{\pi}^{\rm phys.}=140$ MeV. 
 The critical temperature for $ m_q^{\rm crit.} / m_q^{\rm phys.} = 0.08$ is
 $T_c \simeq 170$ MeV. The behavior of $\xi(T)$ for $m_{\pi}(T=0)=30$ MeV 
 (just below the critical value $m_{\pi}^{\rm crit.}$)
 is also shown by the dashed line in Fig.\ref{vevm}(B) for comparison.

\subsection{Chiral limit}
\label{chiral-l}

\begin{figure}[h]
\centerline{
    \epsfysize=6.3cm
    \epsfbox{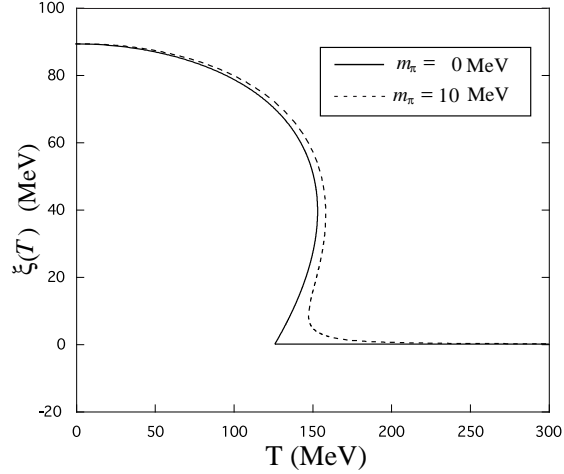}\hspace{0.2cm}
}
     \caption{$\xi(T)$ for $m^{peak}_{\sigma}(T=0) = 550$ MeV 
     with $m_{\pi} =0$ MeV and $m_{\pi}=10$ MeV.}
 \label{chiral}
\end{figure}

 In Fig.\ref{chiral} the chiral condensates are shown for 
 the chiral limit ($m_{\pi}=0$ MeV) and for $m_{\pi}=10$ MeV. 
 The phase transition looks like a first order in these cases. 
 The existence of the multiple solutions of the gap equation
 for the $O(4)$ $\sigma$ model in the mean-field approach 
 has been known for a long time \cite{first}. Our analyses 
 confirm this feature within the framework of OPT.

 However, this first order nature is likely to be
 an artifact of the mean-field approach as discussed 
 in the second reference in \cite{first}:
 the higher loops of massless $\pi$ and almost massless
 $\sigma$ are not negligible near the critical point, and they 
 could easily change the order of the transition.
 In fact, the renormalization group analyses as well as
 the direct numerical simulation on the lattice
 indicate that the $O(4)$ $\sigma$ model has a second order
 phase transition \cite{kanaya}.

 We have also studied the free energy as a function of $T$
 near the chiral limit and found that it has a discontinuity
 near the critical point. 
 This is another sign that the first-order nature
 is an artifact of the approximation.
 (Remember that, the free energy must be a continuous
 function of $T$ irrespective of the order of the phase transition.)

\subsection{Spectral function at finite $T$}
\label{spect-example}

\begin{figure}[h]
\centerline{
    \epsfysize=6.3cm
    \epsfbox{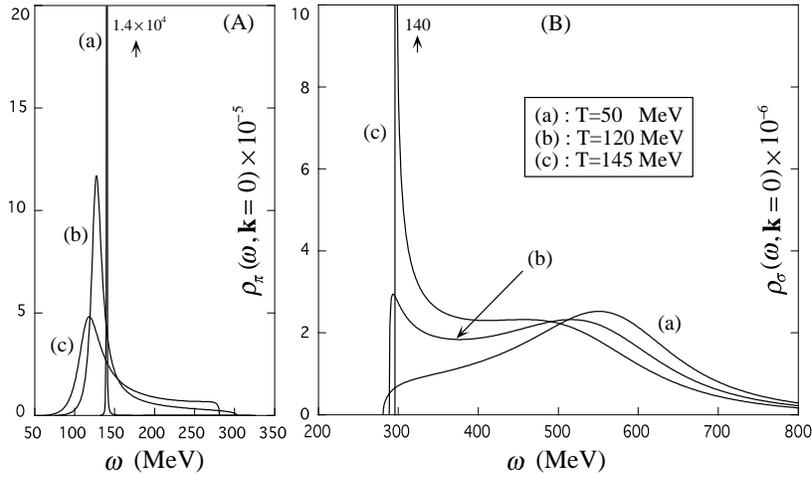}
}
     \caption{Spectral function in the 
     $\pi$ channel (A) and in the $\sigma$ channel
     (B) for $T=50, 120, 145 $ MeV with $m^{peak}_{\sigma} (T=0)=550$ MeV.}
 \label{spectT}
\end{figure}

As one of the non-trivial applications of OPT, 
 we show, in Fig.\ref{spectT}, the spectral functions of $\pi$ and $\sigma$
 at finite $T$ defined in (\ref{spectral}) .

 The figure shows that the spectral function of $\sigma$, 
 which does not show a clear resonance at $T=0$, develops
 a sharp enhancement near 2$\pi$ threshold as $T$ approaches
 $T_c$. This is due to a combined effect of the partial 
 restoration of chiral symmetry and the strong $\sigma - 2 \pi$ coupling. 
 This is an typical example of the softening (or the
 precursor of the critical fluctuation) associated
 with the partial restoration of chiral symmetry  \cite{hk85}.
 The experimental relevance of this softening has 
 been examined in the context of the low-mass diphoton production 
\cite{our}.
 
 In  the $\pi$-channel, a continuum develops by 
 the scattering with thermal pions in the heat-bath;
 $\pi + \pi^{\rm thermal} \rightarrow \sigma $. Because of this 
 process, the pion acquires a width $ \sim 50 {\rm MeV}$
 at $T= 145$ MeV, while the peak position does not show 
 appreciable modification. They are in accordance with
 the Nambu-Goldstone nature of the pion.

\section{Summary}

 We have shown that the optimized perturbation theory (OPT) 
 developed in \cite{our} naturally cures the problems of the
 naive loop-expansion at finite $T$, namely, the breakdown of 
 the naive perturbation at $T \gg T_c$ (due to the hard thermal loops) 
 as well as at $T < T_c$ (due to the tachyonic poles).  

 Furthermore,  OPT has several advantages over  
 other resummation methods proposed so far:

 First of all, the renormalization of the UV divergences, 
 which is not a trivial issue in other methods, 
 can be carried out systematically in  the loop-expansion in OPT. 
 This is because one can separate the self-consistent procedure 
 (Step 3 in Sec. \ref{Newresum}) from the renormalization procedure 
 (Step 2 in Sec. \ref{Newresum}).

 Secondly, the Nambu-Goldstone (NG) theorem is fulfilled  in any give order
 of the loop-expansion in OPT for arbitrary $N$ in $O(N)$ $\phi^4$ theory.
 This is because OPT preserves the global symmetry of the effective potential 
 in each order of the perturbation.

 There are many directions where OPT at finite $T$ may be applied.
 The phase transition in $O(4)$ $\phi^4$ theory as well as 
 the {\em dynamical} critical phenomena near the critical point 
 are one of the most interesting problems to be examined further.  
 The PMS condition for the effective action will be suitable for this purpose.
 OPT may also have relevance to develop an improved perturbation
 theory for gauge theories in which the  weak coupling expansion 
 is known to break down in high orders \cite{linde80}.

\section*{Acknowledgments}

 This work was partially supported by the Grants-in-Aid of 
 the Japanese Ministry of Education, Science and Culture (No. 06102004).
 S. C. would like to thank the Japan Society of Promotion of Science (JSPS)
 for financial support. 

\onecolumn

%
%

\end{document}